\documentclass[useAMS,usenatbib]{mn2e}
\usepackage{graphicx} 
\usepackage{lscape} 
\usepackage{longtable} 
\usepackage{afterpage} 
\usepackage{multirow} 
\usepackage{definitions}
\begin{document}         

\title[L dwarfs in the Hyades]{L dwarfs in the Hyades}

\author[E. Hogan et al.]{E. Hogan\textsuperscript{1}, R.~F. 
Jameson\textsuperscript{1}, S.~L. Casewell\textsuperscript{1}, S.~L. 
Osbourne\textsuperscript{1} and N.~C. Hambly\textsuperscript{2}\\
\textsuperscript{1}Department of Physics and Astronomy, University of 
Leicester, University Road, Leicester, LE1 7RH, UK\\
\textsuperscript{2}Scottish Universities' Physics Alliance (SUPA), Institute 
for Astronomy, School of Physics, University of Edinburgh,\\
Royal Observatory, Blackford Hill, Edinburgh, EH9 3HJ}

\maketitle

\begin{abstract}
\label{abstract}
We present the results of a proper motion survey of the Hyades to search for 
brown dwarfs, based on UKIDSS and 2MASS data. This survey covers 
$\sim275$~deg$^{2}$ to a depth of $K\sim15$~mag, equivalent to a mass of 
$\sim0.05\,M_{\odot}$ assuming a cluster age of $625$~Myr. The discovery of 12 
L dwarf Hyades members is reported. These members are also brown dwarfs, with 
masses between $0.05<M<0.075\,M_{\odot}$. A high proportion of these L dwarfs 
appear to be photometric binaries. 
\end{abstract}


\section{Introduction}
\label{introduction}
The Hyades is the closest cluster to the Sun at a distance of $46.3$~pc 
\citep{pbl1998}, making it an ideal cluster to search for faint, low mass 
objects such as brown dwarfs. However, due to its proximity, the Hyades covers 
a large area of the sky, requiring considerable amounts of telescope time to 
survey the entire cluster. In addition, there is a deficiency of very low mass 
objects in the Hyades \citep{grm1999}, which \citet{bkm2008} suggest is caused 
by the preferential evaporation of the lower mass cluster members.

\begin{figure*}
\begin{center}
\mbox{\includegraphics[angle=270,scale=0.664]{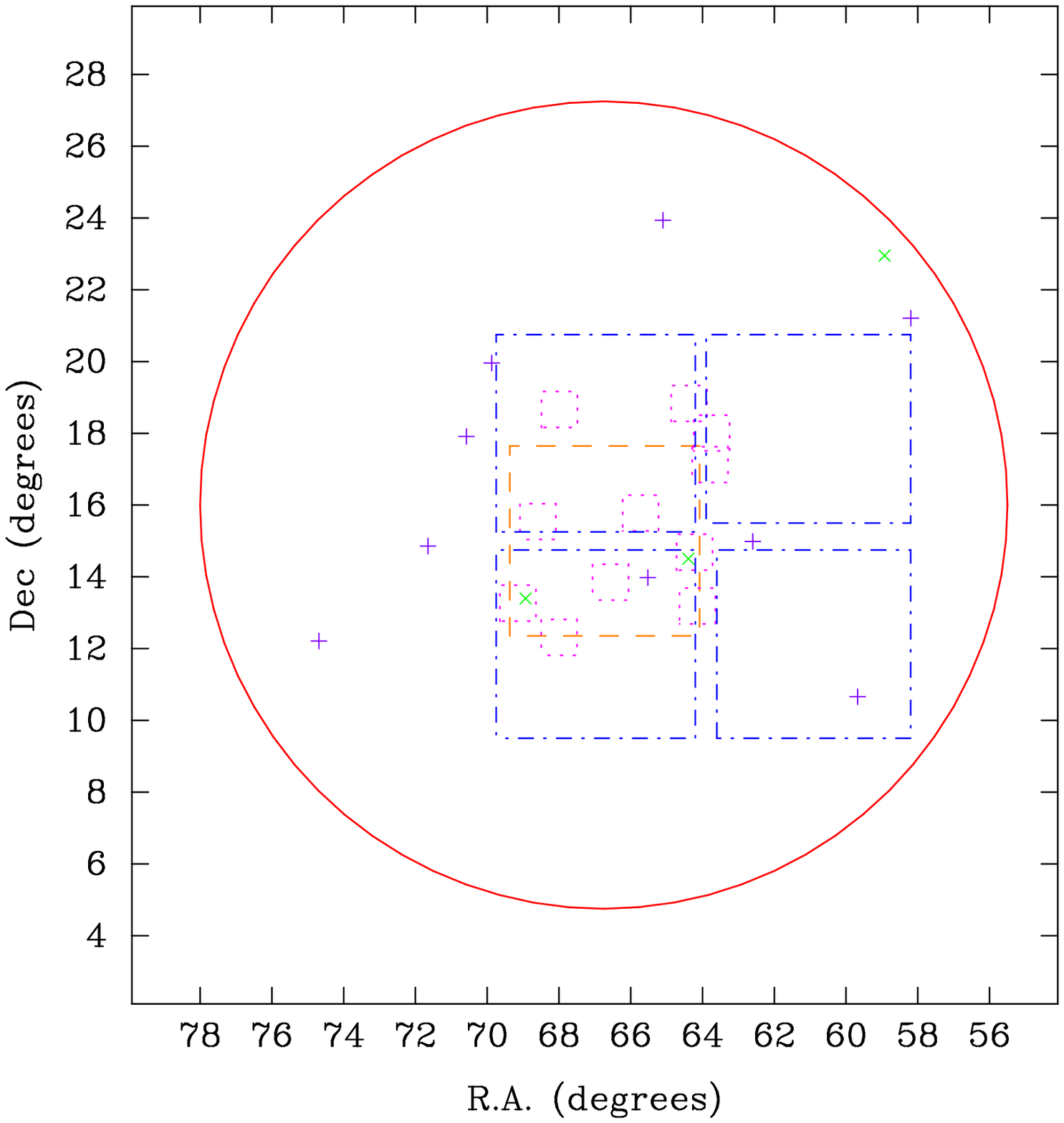}}
\caption{The 11 small magenta dotted squares outline the regions covered by \citet{dkj2002}. The region surveyed by \citet*{grm1999} is indicated by the single orange dashed square. The 4 blue dotted--dashed squares show the areas covered by the survey of \citet{r1992}. The area covered by UKIDSS is indicated by the large red solid circle. The brown dwarf candidates are shown as purple plus symbols and green crosses, where the green crosses indicate the candidates with $M_K>11$~mag.}
\label{previoussurveys}
\end{center}
\end{figure*}

The details of the most recent surveys of the Hyades are shown in 
Figure~\ref{previoussurveys} and Table~\ref{hyadessurveys}. Many of these 
surveys have been specifically designed to be sensitive enough to detect brown 
dwarfs. However, only 2 brown dwarfs have been found so far \citep{bkm2008}. 
\citet{bj2007} used proper motions and the moving group method to find brown 
dwarfs that appear to belong to the Hyades moving group, with the possibility 
that they could be also escaped cluster members. \citet{omb2007} measured 
radial velocities for five of the \citet{bj2007} objects and confirmed that 
they have velocities lying in the $2\sigma$ ellipsoid of the Hyades moving 
group. However, only one T7.5 dwarf, 2MASS~J$12171110-0311131$ (2MJ$1217-03$; 
\citealt{bkb1999}), has a space velocity very close to that of the Hyades 
cluster, making it likely to be an escaped cluster member. The other four 
\citet{bj2007} objects with radial velocity measurements are probably members 
of the Hyades moving group but not escaped cluster members, so they may not be 
exactly coeval with the Hyades cluster. 

Since the Hyades is located in front of the Taurus dark cloud star forming 
region, it is dangerous to select members based purely on photometric 
criteria. Young, more distant objects can be readily confused with genuine 
Hyades members. As Hyades members have a substantial proper motion, candidates 
are selected if they have the correct photometric properties \textit{and} 
proper motions.

This paper presents the results of a survey based on the UKIRT Infrared Deep 
Sky Survey (UKIDSS; \citealt{lwa2007}) and the Two Micron All Sky Survey 
(2MASS; \citealt{scs2006}). This survey covers $\sim275$~deg$^{2}$ of the 
Hyades, corresponding to $\sim52\%$ of the total area of the cluster, assuming 
an area of $\pi r^{2}_{\rm{tidal}}$, where $r_{\rm{tidal}}$ is the tidal 
radius ($10.5$~pc; \citealt{pbl1998}). Members are determined by selecting 
objects with suitable photometric properties and proper motions. Two ``control 
clusters'', both within the $\sim275$~deg$^{2}$ of the survey, are also 
considered to check for contaminants that may coincidentally have photometric 
properties and proper motions similar to that expected of Hyades members. A 
complete survey of all members will not be attempted, since suitably accurate 
photometry, used to separate the M dwarf cluster stars from the M dwarf field 
stars, does not currently exist. The completed UKIDSS Galactic Cluster Survey 
(GCS) of the Hyades will contain $Z$, $Y$, $J$, $H$, and $K$ magnitudes and 
second epoch $K$ measurements. When complete, the GCS will provide adequate 
photometric measurements and proper motions of all objects in the region of 
the Hyades, allowing a complete census of members and an accurate 
determination of the Present Day Mass Function (PDMF) of the Hyades.

\section{The Survey}
\label{survey}
The GCS sub--survey of UKIDSS has so far covered $\sim~275$~deg$^{2}$ of the 
Hyades cluster in the $K$~band to a magnitude of $18.2$~mag, corresponding to 
a detection limit of $5\sigma$. The UKIDSS data was acquired between August 
2005 and March 2007, $5-10$~years after 2MASS covered the same area of sky in 
$J$, $H$ and $K$ to a magnitude, $K=14.3$~mag, corresponding to a detection 
limit of $10\sigma$. The average epoch difference between the UKIDSS and the 
2MASS data is $\sim7.4$~years. \begin{table}
\begin{center}
\caption{Previous surveys of the Hyades\label{hyadessurveys}}
\begin{tabular}{lrr@{ = }lc}
\hline
\hline
\multicolumn{1}{c}{Survey} & \multicolumn{1}{c}{Area} & \multicolumn{2}{c}{Mag Limit} & \multicolumn{1}{c}{Mass Limit}\\
& \multicolumn{1}{c}{[deg$^2$]} & \multicolumn{2}{c}{[mag]} & \multicolumn{1}{c}{[$M_{\odot}$]}\\
\hline
\citet{bkm2008} & 16.0 & $\,\,I$ & $23$ & 0.05\\
& & $\,\,Z$ & $22.5$ &\\
\citet{dkj2002} & 10.5 & $\,\,I$ & $20.3$ & 0.06\\
& & $\,\,Z$ & $19.5$ &\\
\citet{grm1999} & 28.0 & $\,\,J$ & $16.3$ & 0.06\\
& & $\,\,H$ & $15.4$ &\\
& & $\,\,K$ & $14.8$ &\\
\citet{r1992} & 112.0 & $\,\,V$ & $19.5$ & 0.11\\
\hline
\end{tabular}
\end{center}
\end{table}
 The moderate proper 
motions of Hyades members, which range between $74-140$~mas~yr$^{-1}$ 
\citep{bhj1994}, combined with the epoch difference of $5-10$~years, provide 
measureable proper motions between the two surveys. The astrometric accuracy 
of 2MASS for sources between the magnitude range $9<K<14$~mag is 
$\sim70-80$~mas globally. However, multiply-detected sources that fall in the 
overlap region between 2MASS tiles can be independently detected and measured 
to within $\sim40-50$~mas. Since the UKIDSS data is astrometrically calibrated 
using 2MASS objects \citep{lwa2007}, the difference in the position of the 
objects between the UKIDSS and the 2MASS data provides differential proper 
motion measurements. The error on these measurements can be estimated by 
combining the error on the position of the objects in both catalogues with the 
average epoch difference, leading to a typical error of 
$50/7.4\sim7$~mas~yr$^{-1}$. Since the UKIDSS data saturates at $K\sim10$~mag, 
the combined survey covers the range $10<K<15$~mag, making this survey the 
largest, deep proper motion survey of the Hyades to date. \citet{grm1999} used 
2MASS and a POSSII plate to survey the central $28$~deg$^{2}$ of the Hyades 
(Figure~\ref{previoussurveys}) but did not measure proper motions as the epoch 
difference between 2MASS and the POSSII plate was insufficient.

\begin{figure*}
\begin{center}
\mbox{\includegraphics[angle=270,scale=0.664]{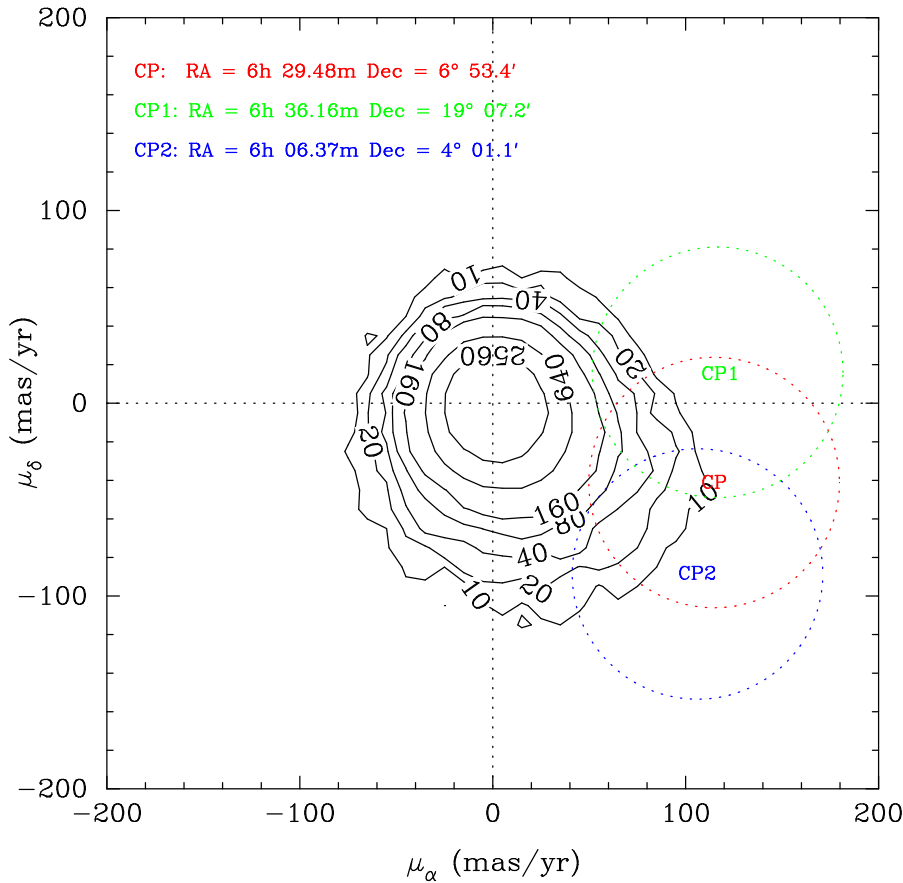}}
\caption{Contour plot showing the proper motions, relative to the Sun, of all the stars in the region of the Hyades, with $J-K>0.6$~mag. The proper motions are calculated as the difference in the positions of the stars matched between the UKIDSS and the 2MASS catalogues. The contours indicate the number of stars per $100$~mas$^{2}$ in proper motion space and show that the distribution of proper motions is not symmetrical about 0,0. The range of proper motions for the Hyades (CP), control cluster 1 (CP1) and control cluster 2 (CP2) are indicated by the dotted circles. CP, CP1 and CP2 are the convergent points for the Hyades, control cluster 1 and control cluster 2, respectively, and are described in the text.}
\label{propermotion}
\end{center}
\end{figure*}

\section{Selection of Hyades Members}
\label{selection}
The large range of proper motions of Hyades members means it is not possible 
to use the method of characterising the cluster and background stars by 
Gaussian distributions, e.g., \citet{dh2004}. Instead, the following method is 
applied. 2MASS and UKIDSS stars are first matched using a $2^{\prime\prime}$ 
matching radius. Proper motions are then measured simply as 
$\mu_{\alpha}={\Delta\alpha\,\rm{cos}\delta}/{\Delta{t}}$ and 
$\mu_{\delta}=\Delta\delta/\Delta{t}$, where $\Delta\alpha$ and $\Delta\delta$ 
are the difference between the R.A. and Dec, respectively, of the 2MASS and 
the UKIDSS coordinates, measured in arc seconds, and $\Delta{t}$ is the epoch 
difference measured in years. 

Initially, all stars with proper motions within the range 
$50<\mu_{\alpha}<180$~mas~yr$^{-1}$ and $-160<\mu_{\delta}<80$~mas~yr$^{-1}$ 
are selected from the the stars matched between the UKIDSS and the 2MASS 
catalogues. This ensures complete coverage of the entire range of proper 
motions of members of the Hyades and both control clusters. Even though there 
is some variation in the proper motions of the Hyades members, they all appear 
to move towards a single point known as the ``convergent point'' (CP; 
$\alpha_{\rm{cp}}=6$h~$29.48$m, $\delta_{\rm{cp}}=6^{\circ}~53.4^{\prime}$; 
\citealt*{mdl2002}). The angular distance, D, between the convergent point and 
each star, and the proper motion angle, $\theta$, measured as the angle 
between the line directly north and the line to the convergent point, is 
calculated for each star using spherical trigonometry:
\begin{equation}
\rm{cos}\,D=\rm{sin}\,\delta\:\rm{sin}\,\delta_{cp}+\rm{cos}\,\delta\:\rm{cos}\,\delta_{cp}\:\rm{cos}\,DA
\label{sphericaltrig1}
\end{equation}
\begin{equation}
\rm{cos}\,\theta=\frac{\rm{sin}\,\delta_{cp}-\rm{sin}\,\delta\:\rm{cos}\,D}{\rm{cos}\,\delta\:\rm{sin}\,D}
\label{sphericaltrig2}
\end{equation}
where $\delta$ and $\delta_{\rm{cp}}$ are the Dec of the star and the 
convergent point, respectively, and DA is the difference in R.A. between the 
star and the convergent point.

Cluster members should have a proper motion angle close to $\theta$. Assuming 
a mean Hyades proper motion of $\sim100$~mas~yr$^{-1}$, the typical error on 
the proper motions of $\pm7$~mas~yr$^{-1}$ corresponds to an error on 
$\theta_{obs}$ of tan$^{-1}\frac{7}{100}\sim4^{\circ}$, where $\theta_{obs}$ 
is the observed proper motion angle. A range of $\pm12^{\circ}$ in 
$\theta_{obs}$ corresponds to $\pm3\sigma$. Therefore, the requirement for a 
member to be selected is chosen to be 
$\theta-12^{\circ}<\theta_{obs}<\theta+12^{\circ}$, leading to a survey which 
is $\sim99.7\%$ complete. 

Cluster members must also have proper motions with the correct magnitude. From 
the theory of moving clusters, it can be shown that the distance, $d$, 
measured in pc, of a member is given by
\begin{equation}
d=\frac{v\,sin\,D}{4.74\,\mu}
\end{equation}
where $\mu$ is the proper motion of the member, measured in arc seconds, and 
$v$ is the cluster velocity, which for the Hyades is equal to 
$46.7$~km~s$^{-1}$ \citep{dyr1984,r1992}. The distance of every potential 
member with a $\theta_{obs}$ within the required range is calculated. Then, a 
second membership condition is imposed; $32<d<60$~pc. This range is based on a 
cluster distance of $46.3$~pc and a tidal radius of $10.5$~pc and also 
includes an additional uncertainty of $\pm3.2$~pc to allow for the $7\%$ 
$1\sigma$ uncertainties associated with the proper motion measurements. This 
additional $1\sigma$ error will still lead to a very complete survey since 
very few members are expected to be located at the tidal radius. This second 
membership condition in effect selects the proper motions with the correct 
magnitude. Since the distance to each object is now known, their absolute 
magnitudes can be calculated.

The final step is to place the members passing the direction and distance 
tests into an HR diagram. Using the $K$ magnitude from UKIDSS and including 
the $7\%$ distance errors mentioned above associated with the error on the 
proper motion measurements, the errors on $M_K$ are calculated 
(Table~\ref{hyadesbd} and Figure~\ref{sequence}). 2MASS provides $J$, $H$ and 
$K$ photometry, but UKIDSS $K$ is more accurate near the 2MASS survey limit. 
The errors on the $J$ magnitudes of the objects from 2MASS generally increase 
as the objects get fainter. As a result, the $J$ magnitudes dominate the 
errors on the $J-K$ colours near the 2MASS survey limit 
(Figure~\ref{sequence}). Since the $K$ magnitudes from UKIDSS are on the MKO 
system \citep{tsv2002}, the $J$ magnitudes from 2MASS are converted to the MKO 
system using the transformations supplied by \citet{sl2004}, so that all $J$ 
and $K$ magnitudes are on the MKO system. 

\section{Contamination}
\label{contamination}
Even though Hyades members are selected as those objects with the correct 
direction and magnitude of proper motion and the correct location in the 
$M_K$, $J-K$ colour--magnitude diagram (CMD), it is still possible that the 
sample will be contaminated by field objects. The proper motions, relative to 
the Sun, of all the stars in the region of the Hyades, with $J-K>0.6$~mag, 
shows the distribution of their proper motions is not symmetrical about 0,0 
(Figure~\ref{propermotion}). In order to estimate the contamination of true 
Hyades members due to objects that may coincidentally have photometric 
properties and proper motions similar to that expected of Hyades members, two 
control areas in proper motion space, which are called ``control clusters'', 
are considered. These control clusters have the same spatial location as the 
Hyades but have slightly different proper motions.

The control clusters need to be as close as possible in proper motion space to 
the Hyades to provide a good estimate of the contamination, bearing in mind 
the asymmetric distribution of the proper motions of stars in the region of 
the Hyades (Figure~\ref{propermotion}). Therefore, the convergent points of 
the control clusters are chosen to be $\pm24^{\circ}$ from the direction of 
the convergent point of the Hyades. This ensures that objects cannot be 
selected as candidates of both the Hyades and either control cluster, within 
the allowed uncertainty on $\theta$. Using the same D, calculated from the 
centre of the Hyades to its convergent point, the convergent points of the two 
control clusters are calculated to be $\alpha_{\rm{cp}1}=6$h~$36.16$m, 
$\delta_{\rm{cp}1}=19^{\circ}~07.2^{\prime}$ and 
$\alpha_{\rm{cp}2}=6$h~$06.37$m, $\delta_{\rm{cp}2}=4^{\circ}~01.1^{\prime}$. 
Members of the control clusters are then found by requiring 
$\theta_{c}-12^{\circ}<\theta_{obs}<\theta_{c}+12^{\circ}$, where $\theta_{c}$ 
may be the direction to either control cluster. The two membership conditions 
applied to the Hyades objects are also applied to the control cluster objects, 
i.e., the control clusters are treated exactly like the Hyades. 

\section{Results}
\label{results}
\begin{figure*}
\begin{center}
\mbox{\includegraphics[bb=86 215 576 522,clip,angle=270,scale=0.539]{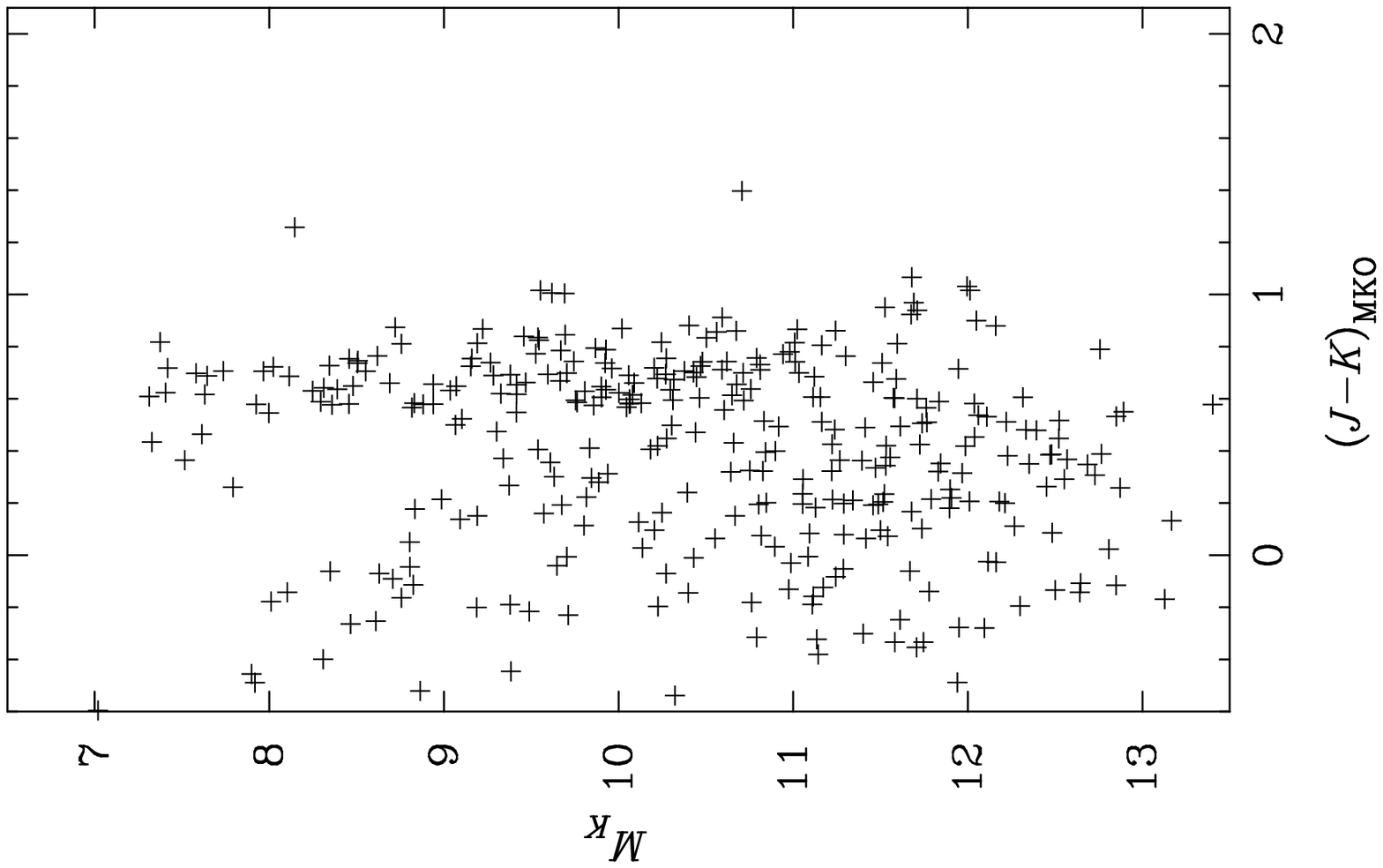}}
\mbox{\includegraphics[bb=86 216 578 523,clip,angle=270,scale=0.539]{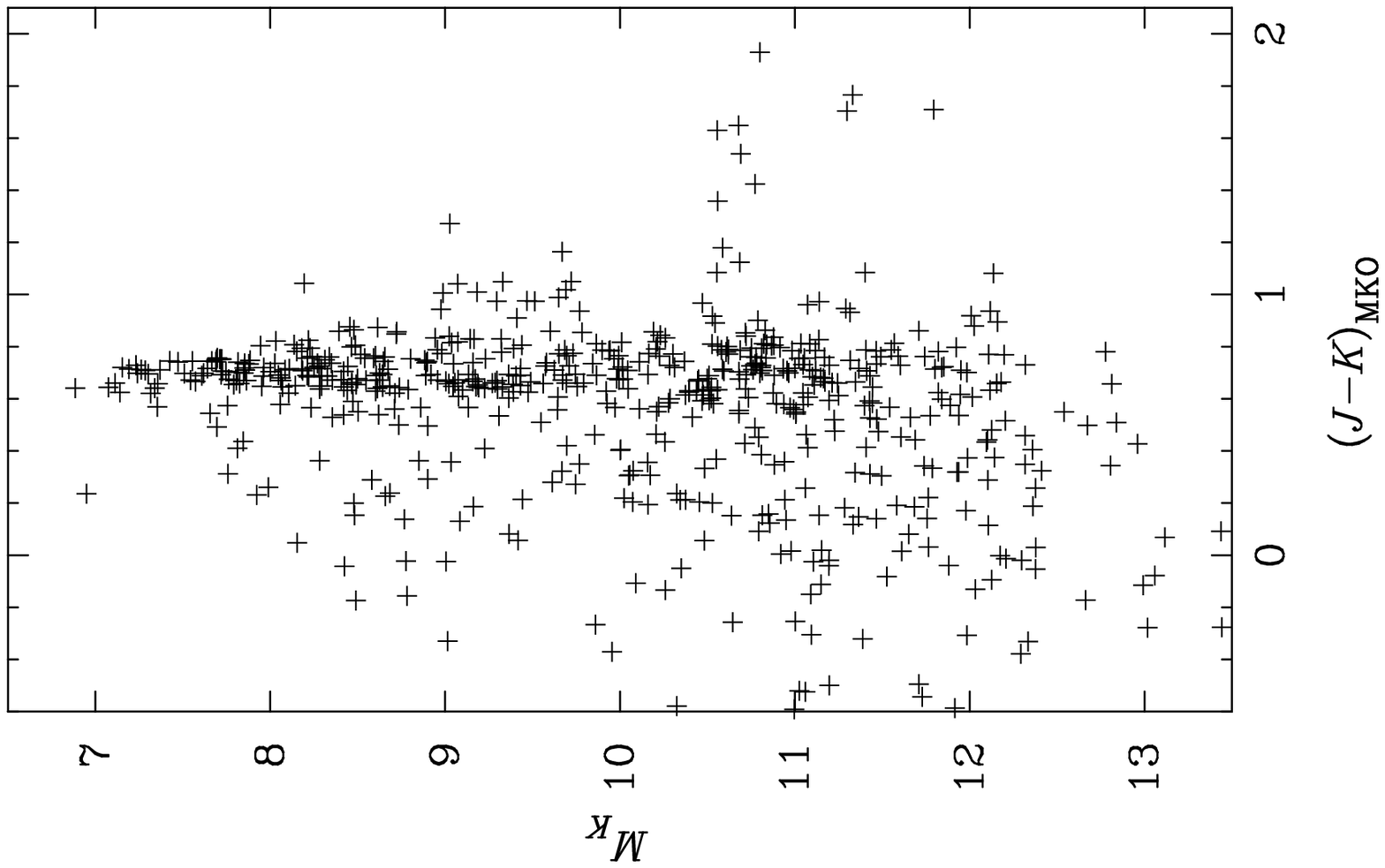}}
\mbox{\includegraphics[bb=86 215 576 522,clip,angle=270,scale=0.539]{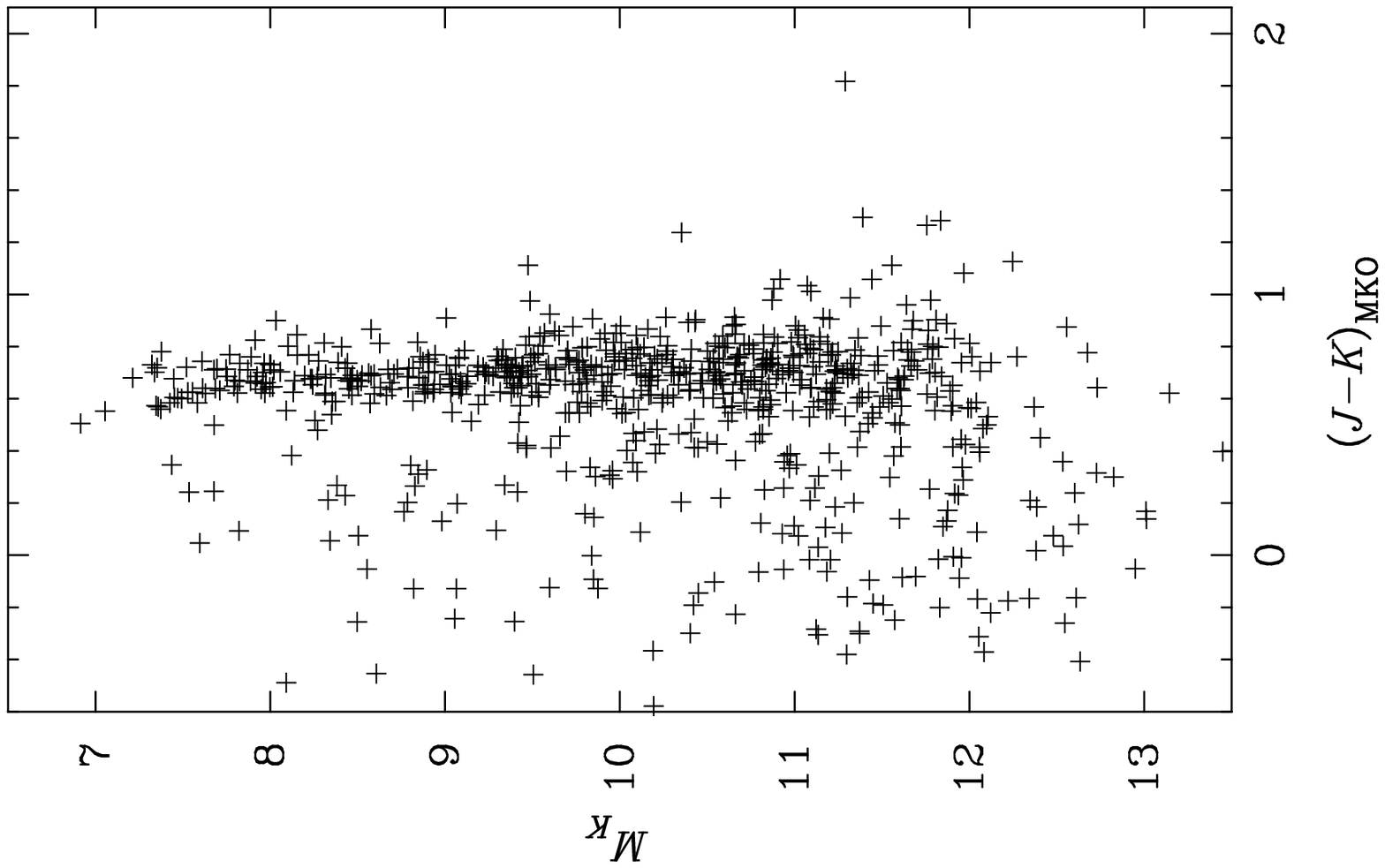}}
\caption{$M_K$, $J-K$ CMD for control cluster 1 (left), the Hyades cluster (centre) and control cluster 2 (right).}
\label{CMD}
\end{center}
\end{figure*}

The $M_K$, $J-K$ CMDs for the Hyades and the two control clusters are shown in 
Figure~\ref{CMD}. It can be seen that control cluster 1 (CP1: 
$\alpha_{\rm{cp}1}=6$h~$36.16$m, $\delta_{\rm{cp}1}=19^{\circ}~07.2^{\prime}$) 
has less M dwarf field stars ($J-K\sim0.7$~mag) than control cluster 2. The 
number of stars in the Hyades M dwarf sequence is an approximate average of 
the number of stars in the M dwarf sequences of the two control clusters. This 
illustrates how difficult it is to separate the Hyades M dwarfs from the field 
M dwarfs. However, the Hyades CMD has a clear sequence of red dwarfs with 
$J-K>1.0$~mag and $10.5<M_K<12.0$~mag. This sequence contains 12 Hyades red 
dwarfs (Table~\ref{hyadesbd}), while each control has 2 red dwarfs with 
similar colours and magnitudes. Since $1.1<J-K<2.0$~mag colours are 
characteristic of L dwarfs, our survey finds $12-2=10$ L dwarfs. However, 
which two are the field contaminants is unknown.

\begin{figure*}
\begin{center}
\mbox{\includegraphics[angle=270,scale=0.664]{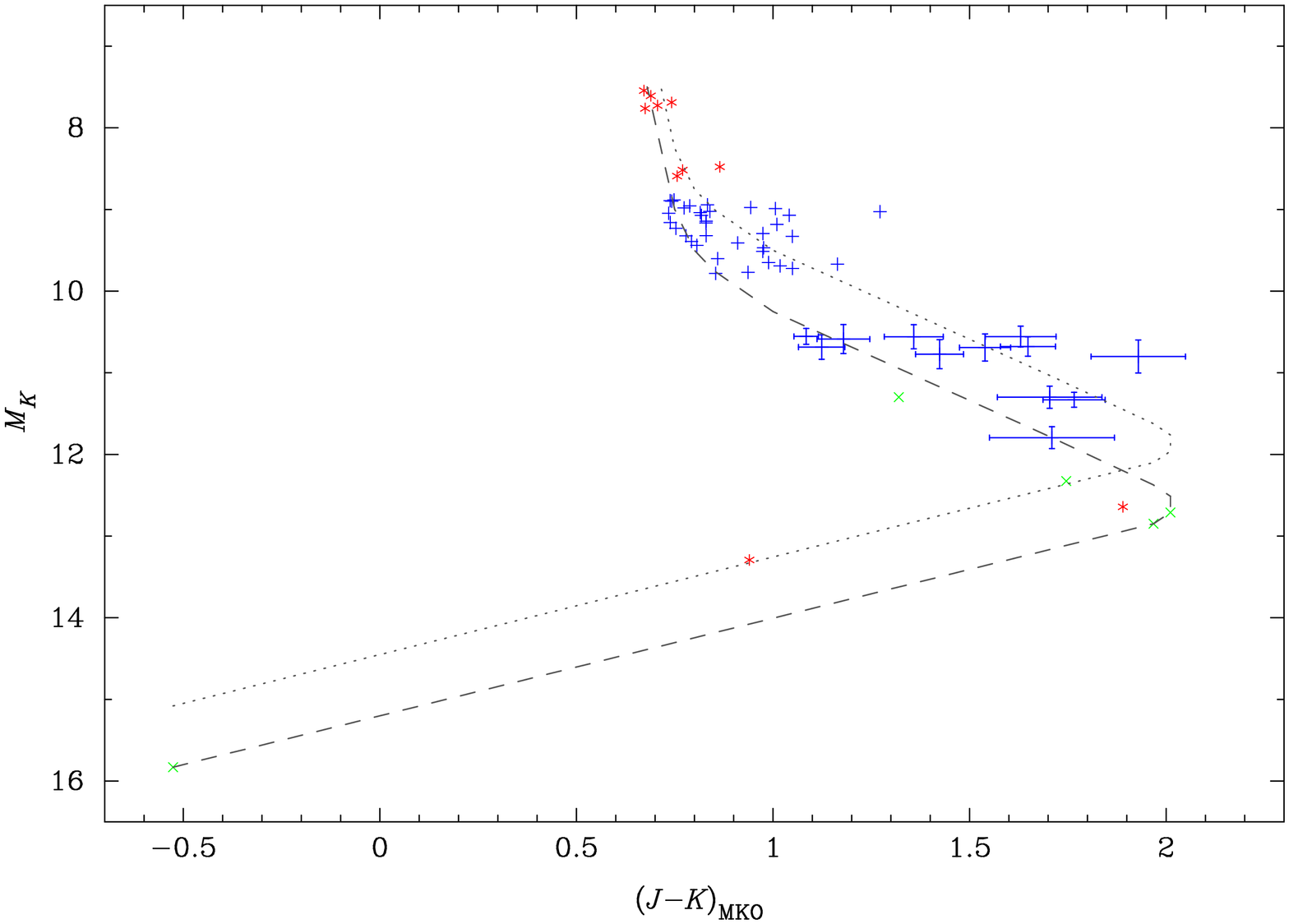}}
\caption{The red stars show the M dwarf Hyades members found by both \citet{bkm2008} and this work, along with the two brown dwarf candidates found by \citet{bkm2008}. The green crosses indicate the members of the Hyades Moving Group (HMG) determined by \citet{bj2007}. The blue plus symbols below $M_K>10.5$~mag with error bars show the 12 brown dwarf candidates discovered in this work, along with the late M dwarf Hyades candidate members. The dashed and dotted lines indicate the single star and equal mass binary sequences, respectively.}
\label{sequence}
\end{center}
\end{figure*}

The Hyades CMD also shows 11 dwarfs with $9<M_K<10$~mag and $1.0<J-K<1.2$~mag, 
whereas control clusters 1 and 2 show 3 and 2 similar dwarfs, respectively. 
Therefore, the Hyades appears to have $12-3=9$ very late M dwarfs with 
$1.0<J-K<1.2$~mag. These are shown in Figure~\ref{sequence} together with 
their redder counterparts, which may be contaminated by field dwarfs. 
\begin{table*}
\begin{center}
\caption{Hyades Brown Dwarfs\label{hyadesbd}}
\begin{tabular}{ccccccccrrccc}
\hline
\hline
& \multicolumn{1}{c}{R.A.} & \multicolumn{1}{c}{Dec} & \multicolumn{1}{c}{$J$} & \multicolumn{1}{c}{$H$} & \multicolumn{1}{c}{$K$} & \multicolumn{1}{c}{$K$} & \multicolumn{1}{c}{$J-K$} & \multicolumn{1}{c}{$\mu_{\alpha}$} & \multicolumn{1}{c}{$\mu_{\delta}$} & \multicolumn{1}{c}{$d$} & \multicolumn{1}{c}{$M_{K}$} & \multicolumn{1}{c}{log~$g$}\\
& \multicolumn{1}{c}{(2000)} & \multicolumn{1}{c}{(2000)} & \multicolumn{1}{c}{(2M)} & \multicolumn{1}{c}{(2M)} & \multicolumn{1}{c}{(2M)} & \multicolumn{1}{c}{(UK)} & \multicolumn{1}{c}{(MKO)} & \multicolumn{2}{c}{[mas/yr]} & \multicolumn{1}{c}{[pc]} & \multicolumn{1}{c}{[mag]} &\\
\hline
1 & 04 20 24.5 & 23 56 13 & 14.60 & 13.85 & 13.42 & 13.35 & 1.08 & 148.98 & -46.41 & 36.2$\,\pm\,$1.63 & 10.55$\,\pm\,$0.097 & 5.27\\
2 & 03 52 46.3 & 21 12 33 & 15.94 & 14.81 & 14.26 & 14.17 & 1.63 & 114.31 & -36.95 & 52.8$\,\pm\,$3.08 & 10.56$\,\pm\,$0.127 &\\
3 & 04 10 24.0 & 14 59 10 & 15.75 & 14.78 & 14.17 & 14.25 & 1.36 & 102.46 & -7.86 & 54.6$\,\pm\,$3.72 & 10.56$\,\pm\,$0.148 &\\
4 & 04 42 18.6 & 17 54 38 & 15.60 & 14.97 & 14.23 & 14.26 & 1.18 & 82.71 & -21.25 & 54.4$\,\pm\,$4.46 & 10.59$\,\pm\,$0.178 & 5.27\\
5 & 03 58 43.0 & 10 39 40 & 15.81 & 14.72 & 14.05 & 14.02 & 1.65 & 125.14 & -24.32 & 46.6$\,\pm\,$2.56 & 10.68$\,\pm\,$0.119 & \\
6 & 04 22 05.2 & 13 58 47 & 15.50 & 14.81 & 14.25 & 14.22 & 1.12 & 99.37 & -23.48 & 50.9$\,\pm\,$3.49 & 10.69$\,\pm\,$0.149 & 5.27\\
7 & 04 39 29.1 & 19 57 35 & 15.99 & 15.03 & 14.36 & 14.31 & 1.54 & 86.98 & -28.14 & 53.0$\,\pm\,$4.06 & 10.69$\,\pm\,$0.166 &\\
8 & 04 58 45.7 & 12 12 34 & 15.60 & 14.55 & 14.02 & 14.03 & 1.42 & 82.99 & -19.34 & 44.8$\,\pm\,$3.68 & 10.77$\,\pm\,$0.178 &\\
9 & 04 46 35.4 & 14 51 26 & 16.68 & 15.29 & 14.52 & 14.61 & 1.93 & 72.04 & -22.32 & 57.8$\,\pm\,$5.36 & 10.80$\,\pm\,$0.202 &\\
10 & 04 17 33.9 & 14 30 15 & 16.54 & 15.43 & 14.84 & 14.70 & 1.70 & 108.27 & -29.45 & 47.8$\,\pm\,$2.98 & 11.30$\,\pm\,$0.136 &\\
11 & 03 55 42.0 & 22 57 01 & 16.11 & 15.05 & 14.28 & 14.21 & 1.77 & 165.09 & -32.33 & 37.6$\,\pm\,$1.56 & 11.33$\,\pm\,$0.091 &\\
12 & 04 35 43.0 & 13 23 45 & 16.73 & 15.77 & 14.80 & 14.88 & 1.71 & 112.04 & -17.86 & 41.4$\,\pm\,$2.56 & 11.79$\,\pm\,$0.134 & 5.13\\
\hline
\end{tabular}
\begin{tabular}{p{0.95\textwidth}}
Columns: 2M is the 2MASS magnitude; UK is the UKIDSS magnitude; $\mu_{\alpha}$ and $\mu_{\delta}$ are the R.A. and Dec components of the proper motion of the brown dwarf, respectively, measured in milli arc seconds per year; $d$ is the calculated distance to the brown dwarf, measured in parsecs; the 2MASS $J$ magnitude was converted to the MKO system before the value of $(J-K)_{\rm{MKO}}$ was determined; log~$g$ is estimated from the DUSTY models \citep{cba2000} only for the brown dwarfs that lie on or near the single star sequence, as shown in Figure~\ref{sequence}.
\end{tabular}
\end{center}
\end{table*}

The comparison of the 2MASS $K$ photometry, converted to the MKO system, with 
the UKIDSS $K$ photometry shows no evidence of variability for the 12 brown 
dwarf candidates within the photometric errors. 

\section{Discussion}
\label{discussion}
The L dwarfs discovered in this survey appear to have a horizontal sequence 
with $M_K\sim10.7$~mag followed by a vertical drop at $J-K\sim1.7$~mag, as 
seen  in the $M_K$, $J-K$ CMD. However, we do not believe this can be the true 
sequence. \citet{jlc2008} have used data from L dwarfs found in Upper 
Scorpius, $\alpha$ Per, the Pleiades, Ursa Major and the Hyades \citep{bj2007} 
to make an empirical age determination for L dwarfs. They find gradients for 
these clusters in the $M_K$, $J-K$ CMD to be 1.95, 1.93, 1.81, 2.88 and 2.14, 
respectively. Ursa Major is clearly discrepant and the average gradient of the 
remainder is 1.96. Such a gradient would fit our L dwarf data if we regarded 
the bluest 3 L dwarfs and the faintest L dwarf to be single L dwarfs. The 
remaining L dwarfs would then have to be binary L dwarfs, which could sit up 
to $0.75$~mag above the single star sequence. These single and equal mass 
binary sequences are shown in Figure~\ref{sequence}. A large number of brown 
dwarf binaries might be expected in the Hyades since tight L dwarf binaries 
are less likely to be lost than single L dwarfs due to their greater system 
mass. There are still 3 L dwarfs that remain above the equal mass binary 
sequence. This could be explained by experimental error, accepting that they 
are the expected contaminants or conceivably triple systems. If the apparent 
$K$, $J-K$ CMD is plotted, its shape is not significantly different to the 
$M_K$, $J-K$ CMD.

The single L dwarf sequence must turn around at $J~-~K~\sim~2.0$~mag, 
encompassing the two reddest \citet{bj2007} L dwarfs and continuing on through 
the T7.5 dwarf 2MJ$1217-03$. The T dwarf CFHT--Hy--21 ($J-K\sim1.9$~mag; 
\citealt{bkm2008}) fits this interpretation quite well. CFHT--Hy--20 
($J-K\sim0.9$~mag), on the other hand, sits $\sim0.75$~mag above the proposed 
LT transition sequence. This object would therefore be a binary system, which 
is consistent with the belief that many LT transition objects are binaries 
\citep{tbk2003}.

The single and binary L dwarf sequences are continued to match the M dwarfs 
found by both \citet{bkm2008} and this survey. Note there is a gap of 
$\sim0.7$~mag between the late M dwarfs and the L dwarfs. This could be 
attributed to the ``missing M dwarfs'', a deficit of M7--M8 dwarfs found in 
other clusters \citep{dpj2002}.

Using the NEXTGEN models \citep{bca1998}, the bottom of the hydrogen burning 
main sequence ($0.075\,M_{\odot}$) occurs at $M_K\sim10.3$~mag, assuming a 
cluster age of $625$~Myr. This almost exactly coincides with the brightest of 
the 12 L dwarfs. Therefore, the 12 L dwarfs are also brown dwarfs. Applying 
the DUSTY models \citep{cba2000}, the masses for the 12 L dwarfs range between 
$0.05<M<0.075\,M_{\odot}$. One of the 12 L dwarfs was previously known (2MASSW 
J0355419+225702; L3; \citealt{krl1999}) but not recognised as a Hyades member.

\section{Conclusions and Future Work}
\label{conclusions}
We have found 12 L dwarfs, which are also brown dwarfs, in our survey of the 
Hyades. The level of contamination by field L dwarfs is estimated to be 
2 out of 12. Deciding the Hyades L dwarf sequence from these 12 objects is not 
straightforward since there seem to be more binary L dwarfs than might be 
expected in a younger cluster. Many of the L dwarfs appear to be unresolved 
binaries. If these binaries could be spatially resolved using adaptive optics, 
they might prove to be a useful resource from which dynamical masses could be 
obtained for brown dwarfs of known age. 

When the UKIDSS GCS is complete with $Z$, $Y$, $J$, $H$ and $K$ magnitudes and 
proper motions, it will be possible to make a survey of the Hyades $\sim3$ 
magnitudes deeper than this survey. The limit of the final survey will be 
$M_K\sim15$~mag and should therefore find fainter L and T dwarfs in the 
Hyades. It will also be possible to make a good census of low mass Hyades 
members and derive a precise Present Day Mass Function of the Hyades. 

\section{Acknowledgements}
\label{acknowledgements}
EH and SLC are Post Doctoral Research Associates, funded by STFC. SLO was a 
PPARC supported Postgraduate Student. This work is based in part on data 
obtained as part of the UKIRT Infrared Deep Sky Survey. This publication makes 
use of data products from the Two Micron All Sky Survey, which is a joint 
project of the University of Massachusetts and the Infrared Processing and 
Analysis Center/California Institute of Technology, funded by the National 
Aeronautics and Space Administration and the National Science Foundation.

\bibliographystyle{mn2e}
\bibliography{bdreference}

\end{document}